# Two source emission behavior of projectile fragments alpha in $^{84}$Kr interactions at around 1 GeV per nucleon

M. K. Singh,[1, 2] Ramji Pathak,[1] and V. Singh[2]

1. Department of Physics, Tilakdhari Postgraduate College, Jaunpur - 222002

2. Nuclear and Astroparticle Physics Laboratory, Department of Physics,

Banaras Hindu University, Varanasi - 221005

**Abstract:**

The emission of projectile fragments alpha has been studied in $^{84}$Kr interactions with nuclei of the nuclear emulsion detector composition at relativistic energy below 2 GeV per nucleon. The angular distribution of projectile fragments alpha in terms of transverse momentum could not be explained by a straight and clean-cut collision geometry hypothesis of Participant – Spectator (PS) Model. Therefore, it is assumed that projectile fragments alpha were produced from two separate sources that belong to the projectile spectator region differing drastically in their temperatures. It has been clearly observed that the emission of projectile fragments alpha are from two different sources. The contribution of projectile fragments alpha from contact layer or hot source is a few percent of the total emission of projectile fragments alphas. Most of the projectile fragments alphas are emitted from the cold source.

## I. Introduction

Nuclear fragmentation is an important experimental phenomenon in nucleus – nucleus collisions at relativistic high energy [1]. The emission of projectile fragments alpha has been studied in several experiments and the general consensus was that they are coming out from the projectile spectator part in the context of a PS Model [2]. Large number of models has been introduced in the investigation of projectile fragmentation in the nucleus – nucleus interactions at very high energies but at just relativistic kinetic energy (below few GeV), only a few models have been found suitable [3]. The PS Model is the simple and basic model for the study of the high energy nucleus – nucleus collisions. According to the PS Model, the interacting system in high energy nucleus – nucleus collisions can be divided into three parts: target and projectile spectators and a participant region. The overlapping part of two colliding nuclei is called the participant region and the other parts are called the target and the projectile spectators, respectively. In this paper, we mainly focus on projectile spectator region. Earlier data from lighter projectiles has been successfully explained with the help of PS Model but data from high energy heavy ion interactions pointed finger towards the PS Model because experiments observed many projectile fragments alpha having larger angle of emission that could not be explained by the straight and clean-cut collision geometry picture of the PS Model. Therefore, this large angle scattering indicated the failure of the PS Model and forced us to rethink and modify this model according to the physics requirements and the idea of multiple sources i.e. emission of projectile fragments alpha from different sources having different temperature has been introduced [4].

In this paper, we analyzed the multiplicity and transverse momentum distributions of projectile fragments alpha produced in $^{84}$Kr interactions with the nuclei of the

Nuclear Emulsion Detector's (NED) target at around 1 GeV per nucleon kinetic energy [5]. The NED is having very high position and angular resolution and its composition is acting as target for the projectile to interact, so it is widely used in the investigation of nuclear fragmentation [6]. The NED which provides $4\pi$ geometrical coverage, therefore, allows an exclusive type of analysis on an event-by-event basis with detailed information about the fragmentation mechanism of nucleus – nucleus interactions. Experimental details are described in section 3 while required modification in PS Model has been discussed in section 2. Observed results, which explain data of projectile fragmentation using a two source PS Model, are discussed in section 4.

## II. The Model

It is strongly believed that the nuclear collision geometry plays an important role in the study of particular behavior of the nucleus - nucleus collisions. According to the PS Model [2], the overlapping part of two colliding nuclei is called the *participant*, from where newly created and / or freshly produced particles occurs and the remaining parts of nuclei which do not participate in the collision are called the *target spectator* and the *projectile spectator*, respectively. In the collisions, due to the existence of the relative motion between the participant and the spectator, the friction is assumed to be caused on the contact layer. In this situation, both the participant and the spectator get the heat due to friction. It takes some time when the contact layer transmits the heat of friction to the rest part of the spectator and therefore, we believe, that this may be the cause of temperature gradient in the spectator region of projectile. The contact layer and the rest part, which are separated from each other because of the heat of friction. Therefore, the contact layer and the rest part of the spectator are considered as two sources to emit nuclear fragments with two different temperatures. It could be possible that during the collision contact layer portion have highest temperature after participant region. The fall in temperature is rapid towards the farther side of projectile spectator region. The change in temperature follows exponential decay nature and it could be explained from the measured charge spectrum of the projectile fragments. We can also be possible that the temperature is almost constant in a layer and the thickness of layers increases with distance from the contact layer as shown in figure 1. The two emission sources are the projectile spectator contact layer i.e. hot spectator having comparatively high temperature and the other part i.e. cold spectator having comparatively low temperature. This could lead the whole spectator to a non-equilibrium state, but the contact layer and the rest part to be the local equilibrium state.

Let $n_c$ and $n_o$ denote the multiplicities of projectile fragments alpha emitted from the closer contact layers from the participant region and the rest part of the projectile spectator, respectively. It is also possible that heavy projectile clusters gets excited and decays in a few projectile fragments alpha in a very short time and the distance travelled before decay is actually close to the vertex of the event. The multiplicity of projectile fragments alpha measured in the final state of interaction is denoted by $n_\alpha$. The relationship between the mean multiplicities is as follows

$$<n_c> = k<n_\alpha> \tag{1}$$

and

$$<n_o> = (1-k)<n_\alpha>. \tag{2}$$

Where, k is an extra parameter.

Let $P_c(n_c)$ be the probability for the contact layer to emit $n_c$ fragments, and $P_o(n_o)$ is that for the rest part of the spectator to emit $n_o$ fragments. Then,

$$P_c(n_c) = \frac{1}{<n_c>} \exp(\frac{-n_c}{<n_c>}) / [1 - \exp(\frac{-n_{cm}}{<n_c>})] \qquad (3)$$

and

$$P_c(n_o) = \frac{1}{<n_o>} \exp(\frac{-n_o}{<n_o>}) / [1 - \exp(\frac{-n_{om}}{<n_o>})]. \qquad (4)$$

Where $n_{cm}$ and $n_{om}$ are the maximum values of $n_c$ and $n_o$, respectively.

The probability $P(n_\alpha)$ for the spectator to emit $n_\alpha$ fragments depends only on the sum of $n_c$ and $n_o$. That is

$$P(n_\alpha) = \int_0^{n_\alpha} P_c(n_c) P_0(n_\alpha - n_c) dn_c \qquad (5)$$

$$P(n_\alpha) = \frac{1}{(<n_c><n_o>)} \int_0^{n_\alpha} \exp(\frac{-n_c}{<n_c>}) \exp\{-\frac{(n_\alpha - n_c)}{<n_o>}\} dn_c$$

$$/ [1 - \exp(\frac{-2n_{cm}}{<n_\alpha>})][1 - \exp(\frac{-2n_{om}}{<n_\alpha>})]$$

$$= \frac{4n_\alpha}{<n_\alpha>^2} \frac{\exp(\frac{-2n_\alpha}{<n_\alpha>})}{\{[1 - \exp(\frac{-2n_{cm}}{<n_\alpha>})][1 - \exp(\frac{-2n_{om}}{<n_\alpha>})]\}}. \qquad (6)$$

In the case of $k = 0.5$ and the maximum multiplicity of projectile fragments alpha $n_{\alpha max} \approx 2n_{cm} \approx 2n_{om}$, then

$$P(n_\alpha) \approx \frac{4n_\alpha}{<n_\alpha>^2} \exp(\frac{-2n_\alpha}{<n_\alpha>}) / [1 - \exp(\frac{-n_{\alpha max}}{<n_\alpha>})]^2. \qquad (7)$$

Especially for a maximum projectile fragments alpha ($n_{\alpha max}$) values, above Eq. (7) becomes

$$P(n_\alpha) \approx \frac{4n_\alpha}{<n_\alpha>^2} \exp\left(\frac{-2n_\alpha}{<n_\alpha>}\right). \tag{8}$$

The projectile fragments alpha emission from each source is assumed to be isotropic in the source rest frame. Let z denote the direction of the incident projectile and xoz plane be the reaction plane. The three components of projectile fragments alpha momentum ($p_{xyz}$) in the source rest frame are assumed to have Gaussian distributions with the same width $\sigma_p$. Therefore, we have

$$P_{px,y,z}(p_{x,y,z}) = \frac{1}{(2\pi\sigma_p)^{1/2}} \exp\left(\frac{-p_{x,y,z}^2}{2\sigma_p^2}\right). \tag{9}$$

Then, the transverse momentum

$$p_T = (p_x^2 + p_y^2)^{1/2}, \tag{10}$$

has Rayleigh scattering distribution as

$$P_{pt}(p_T) = \frac{p_T}{\sigma_p^2} \exp\left(\frac{-p_T^2}{2\sigma_p^2}\right). \tag{11}$$

Considering the two-source emission of projectile fragments alpha, the final-state transverse momentum (p$_T$) distribution should be the sum of two Rayleigh scattering distributions, i.e. active sources are represented by each distribution [7]

$$P_{pT}(p_T) = (\frac{A_H p_T}{\sigma_H^2})\exp(\frac{-p_T^2}{2\sigma_H^2})+(\frac{A_L p_T}{\sigma_L^2})\exp(\frac{-p_T^2}{2\sigma_L^2}), \qquad (12)$$

where $\sigma_H$ and $\sigma_L$ are the p$_T$ distribution widths of projectile fragments alpha emitted from the sources of high and low temperatures, respectively. A$_H$ and A$_L$ denotes the number of projectile fragments alpha contribution from the two sources where temperature of the sources are considered as high and low, respectively.

## III. Experimental details

This experiment has been performed in a stack of NIKFI BR-2 Nuclear Emulsion - Detectors (NED) having 600 μm thickness exposed at GSI Darmstadt Germany to a beam of $^{84}$Kr nuclei at around 1 A GeV [8]. The measurements have been done at Banaras Hindu University by the scanning of NED volume, having dimension 9.8 cm × 9.8 cm × 0.06 cm, with the help of Olympus BH-2 transmitted light-binocular microscope under 100X oil emersion objective and 15X eyepieces. There are two standard methods for scanning of the NED that are explained in Ref. [9]. A total of 600 inelastic interactions of $^{84}$Kr nuclei have been picked up and the charge and angle measurements of projectile fragments of these events have been done. Grain, blob and hole density, and delta ray counting measurements have been done for the estimation of the charge of the light projectile fragments with the accuracy of a unit charge and these methods have been explained in detail in Ref. [10]. For angle measurement of the projectile fragment a small but very effective device called "Gonio-meter" has been used having a least count better than a quarter of a degree.

## IV. Results and Discussion

The transverse momentum distributions of projectile fragments alpha produced in the nucleus-nucleus collisions at relativistic energy have been analyzed by using a two emission source of projectile's alpha fragments picture as shown in Figure 1, and each source has drastic change in temperatures. Both alpha emission sources belong to the projectile spectator region and are mainly the projectile spectator's contact layer which is hot due to close contact with the participant region and the other source is the rest part of the projectile spectator which we assumed to be colder than the first one. The hot portion of the projectile spectator with the high excitation degree will emit lighter fragments with high temperature and the momentum distribution of such fragments may tell us about their special distribution into the hot spectator region. The cold portion of the projectile spectator with low excitation degree will generally emit heavy charged fragments with low temperature but can also emit lighter charge fragments too. In the NED based experiment, it is not possible to make direct momentum measurement of charge particles / projectile fragments. However, the transverse momentum can indirectly be calculated by using the fact that the fragments have nearly the same momentum per nucleon as that of the projectile in case of fixed target experiment, because when a projectile nucleus with relativistic energy collides with a target nucleus the projectile fragments emitted retain more or less the same momentum per nucleon as of the projectile nucleon. So if $p_o$ is the momentum of the incident projectile, the transverse momentum of the fragment of charge Z can be calculated by using the following relation:

$$p_T = A_F p_o \, Sin\theta, \qquad (13)$$

where $A_F$ is the mass number of the fragments and $\theta$ is the emission angle of the fragments with respect to the projectile direction. Therefore, in the NED experiment, the pseudo-transverse momentum can be obtained from the measurement of the emission angles.

In order to test the two source fragment emission picture, we compare the sum of two Rayleigh scattering distributions of the transverse momentum with the observed data for similar projectile but with different kinetic energies. The multiplicity distribution of projectile fragments alpha at different energies is shown in Figure 2 and the Gaussian function fitting's parameter for the distribution are tabulated in Table 1. It is clear from the figure and table that the fitted values of the Gaussian function for multiplicity distribution is dependent on the incident energy of the projectile and it also reflects that the maximum number of projectile fragments alpha emitted in an interaction is increasing in both the cases when projectile kinetic energy is high and low. The transverse momentum distributions of projectile fragments alpha at different kinetic energies are shown in the Figure 3. It can be seen from the figure 3 that the tail portion of the $p_T$ distribution have one extra shape with peak at around 480 MeV/c and less than 5% of total projectile fragments alpha are contributing in this distribution. The transverse momentum distribution of projectile fragments alpha has been plotted for different kinetic energies and a solid curve is the sum of two Rayleigh scattering distribution function as described in Eq. (12) are superimposed in Figure 4 and the values of the fitted parameters at different energies are tabulated in table 2. From this table we can see that the values of $\sigma_H$ and $\sigma_L$ are decreasing with decrease in incident energies of the projectiles. It reflects that as the kinetic energy of projectile will be less and less the hot and cold regions must be smaller and smaller i.e. the thermal equilibrium can be achieved very quickly at low kinetic energy of the projectile. In the frame of the Maxwell's ideal gas model, the corresponding temperatures are also derived for both the sources and tabulated in Table 2. Therefore it is clear that relativistic fragments produced in high energy nucleus-nucleus collisions can be regarded as the result of a two-source emission. We can conclude that two source model gives a reasonable description in the case of projectile fragments alpha emitted in $^{84}$Kr nuclei interactions with target nuclei at different kinetic energies below 2 GeV per nucleon.

## V. Conclusions

It is very interesting to study the projectile fragmentation of heavy ions such as $^{84}$Kr as some of the fragmentation characteristics that does not show strong dependence on projectile kinetic energy but have strong dependence on the mass number of the projectile. The idea of projectile fragments alpha emissions being from two different sources of projectile spectator. The main conclusions drawn from this experimental study are as following.

The emission probability of single projectile fragment alpha in an interaction is gradually decreasing with projectile kinetic energy that reflects that the multiple projectile fragments alpha have more chance of emission during interaction keeping the average projectile fragments alpha value almost unchanged.

Our observations confirm the idea that the relativistic projectile fragments alpha produced in high energy heavy ion collisions can be regarded as the result of a two source emission. The transverse momentum distributions of relativistic fragments can be described by two-source emission picture. The distribution of transverse momentum is the sum of two Rayleigh distributions. Generally both sources belong to the projectile spectator part. One projectile fragments alpha emission source is very close to the participant region of the collision and acts as a hot source with high temperature and projectile fragments alpha belonging to this part are distributed in the tail portion of the transverse momentum distribution of projectile fragments alpha. The number of such type of projectile fragments alpha is a few percent of the total projectile fragments alpha. The other source of projectile fragments alpha emission is far from the participant region and has very low temperature and, we believe that the change in temperature in this part is sharp and follows an exponential law. Most of the emitted projectile fragments are from this region of the projectile spectator. As the projectile kinetic energy becomes less and less the area or volume of the rest part becomes larger and larger and play an important role of heavy fragment emission.

The derived values of corresponding temperatures in the frame of the Maxwell's ideal gas model have shown change in temperature in both the sources with projectile kinetic energy.

## Acknowledgements

Authors are thankful to the accelerator division staffs at GSI/SIS and JINR (Dubna) for exposing nuclear emulsion detector with $^{84}$Kr beam and their other helps.

**Table 1:** Multiplicity distribution of projectile fragment alpha has been fitted with the Gaussian function and the fitting parameters values are tabulated in this table.

| Projectile | Energy (A GeV) | Constant | Mean | Sigma |
|---|---|---|---|---|
| $^{84}$Kr | 1.7 | 23.00 ± 0.67 | 1.49 ± 0.28 | 2.79 ± 0.22 |
| $^{84}$Kr | 1.5 | 27.34 ± 0.39 | 1.30 ± 0.94 | 2.64 ± 0.39 |
| $^{84}$Kr | 0.95-0.80 | 22.79 ± 0.56 | 1.37 ± 0.85 | 2.77 ± 0.55 |
| $^{84}$Kr | 0.80-0.50 | 21.95 ± 0.92 | 1.38 ± 0.16 | 2.97 ± 0.39 |
| $^{84}$Kr | 0.50-0.08 | 19.84 ± 0.58 | 2.29 ± 0.19 | 2.68 ± 0.18 |

**Table 2:** Rayleigh scattering function's fitting parameters for the concept of two sources of projectile fragments alpha emission during nucleus – nucleus interactions at high energy. Rayleigh scattering function was fitted on the observed data of the transverse momentum distribution of $^{84}$Kr interactions with target nuclei at different kinetic energies ranging from around 2 to 0.5 GeV per nucleon.

| Projectile | Energy (A GeV) | $A_H$ | $A_L$ | $\sigma_H$ (MeV/c) | $T_H$ (MeV) | $\sigma_L$ (MeV/c) | $T_L$ (MeV) |
|---|---|---|---|---|---|---|---|
| $^{84}$Kr | 1.7 | 0.5 | 0.5 | 175 | 8.16 | 96 | 2.46 |
| $^{84}$Kr | 1.5 | 0.5 | 0.5 | 172 | 8.02 | 95 | 2.43 |
| $^{84}$Kr | 0.95 – 0.80 | 0.5 | 0.5 | 170 | 7.93 | 93 | 2.38 |
| $^{84}$Kr | 0.80 – 0.50 | 0.5 | 0.5 | 169 | 7.88 | 92 | 2.36 |

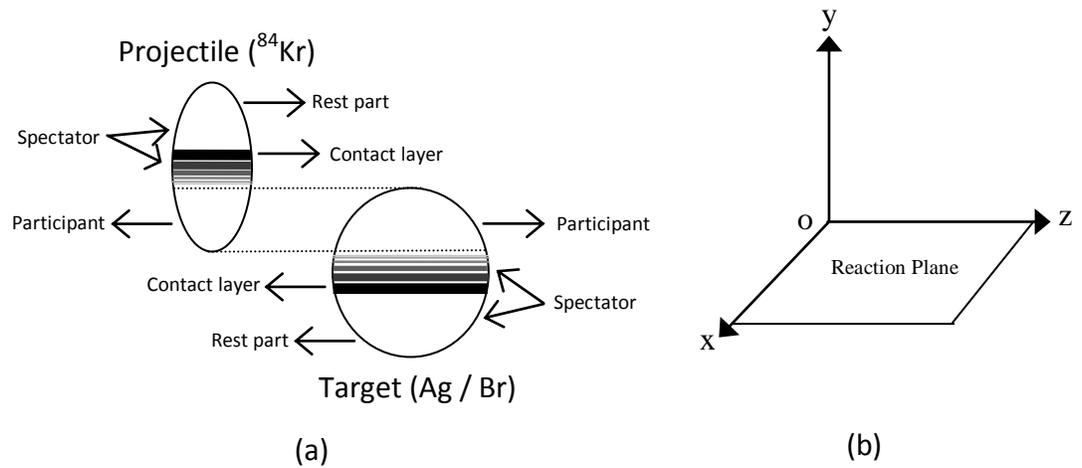

**Fig. 1:** (a) Schematic overview of the contact layer including fine layers of temperatures and the other part of the projectile and target spectators across the incoming direction of the projectile. Darker lines representing lower temperature regions. The projectile is approaching toward the target and target is at rest. Here we are considering only two basic regions of the projectile spectator i.e. contact layer very close to the participant region and rest of the spectator part just for the convenience in calculation. Schematic diagram of the coordinate system of interaction geometry in the laboratory frame has been shown along with reaction plane of the interaction in figure (b).

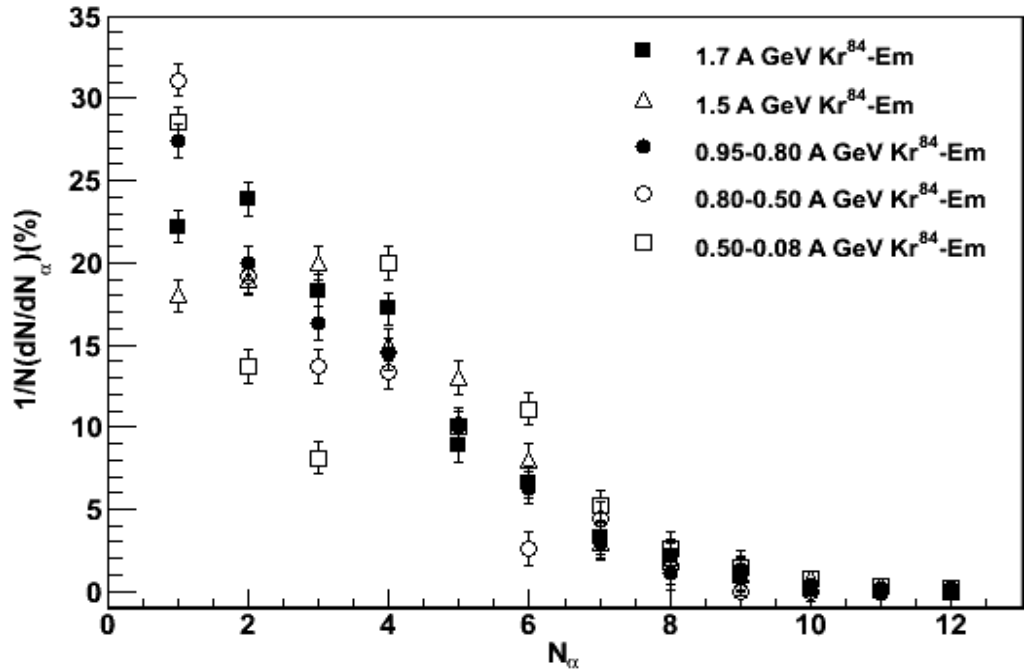

**Fig. 2:** Multiplicity distribution of projectile fragment alpha emitted in $^{84}$Kr nuclei interactions with emulsion target nuclei at different energy.

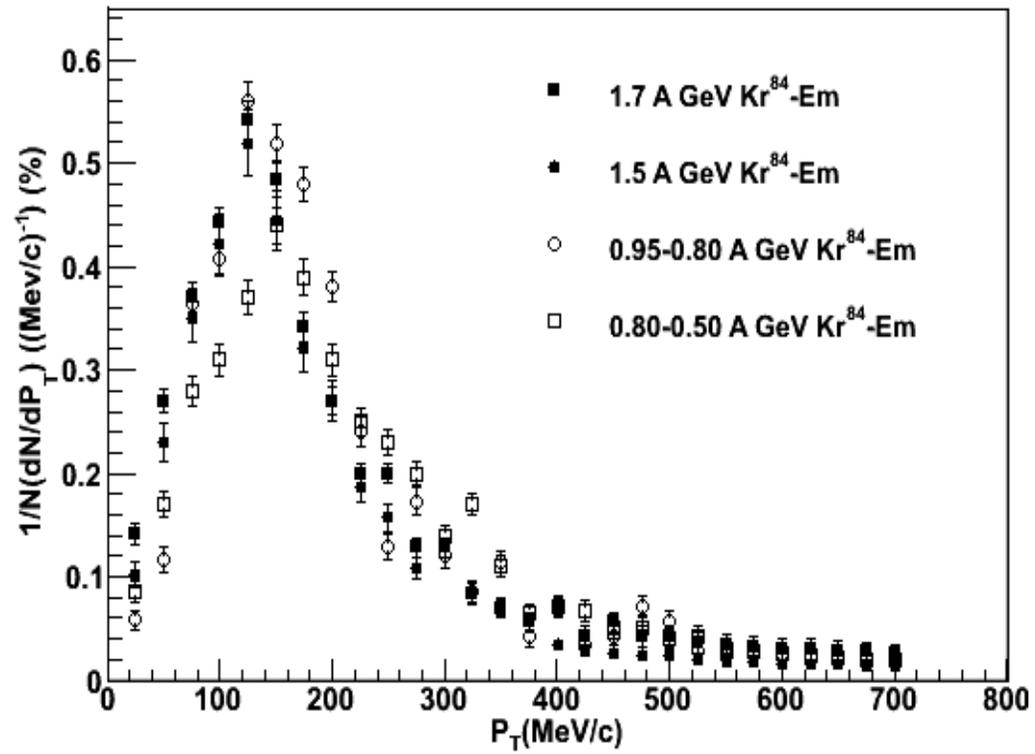

**Fig. 3:** Transverse momentum distribution of projectile fragments alpha emitted in $^{84}$Kr nuclei interactions with emulsion target nuclei at different kinetic energies.

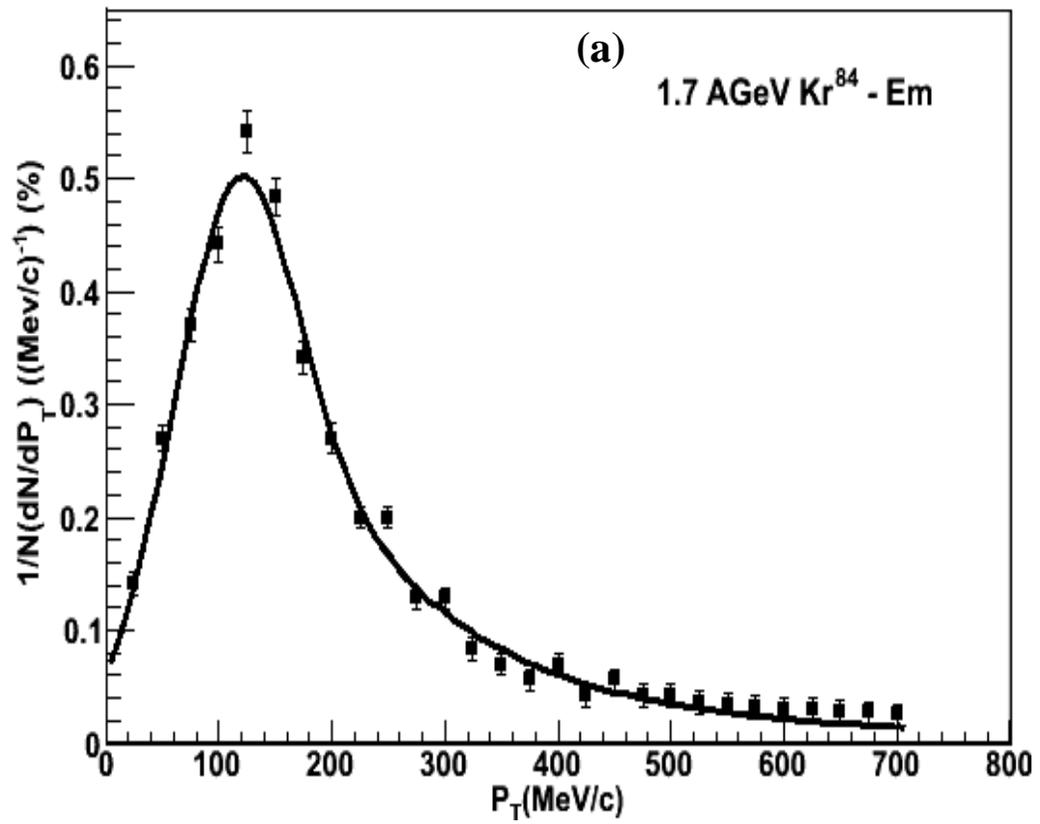

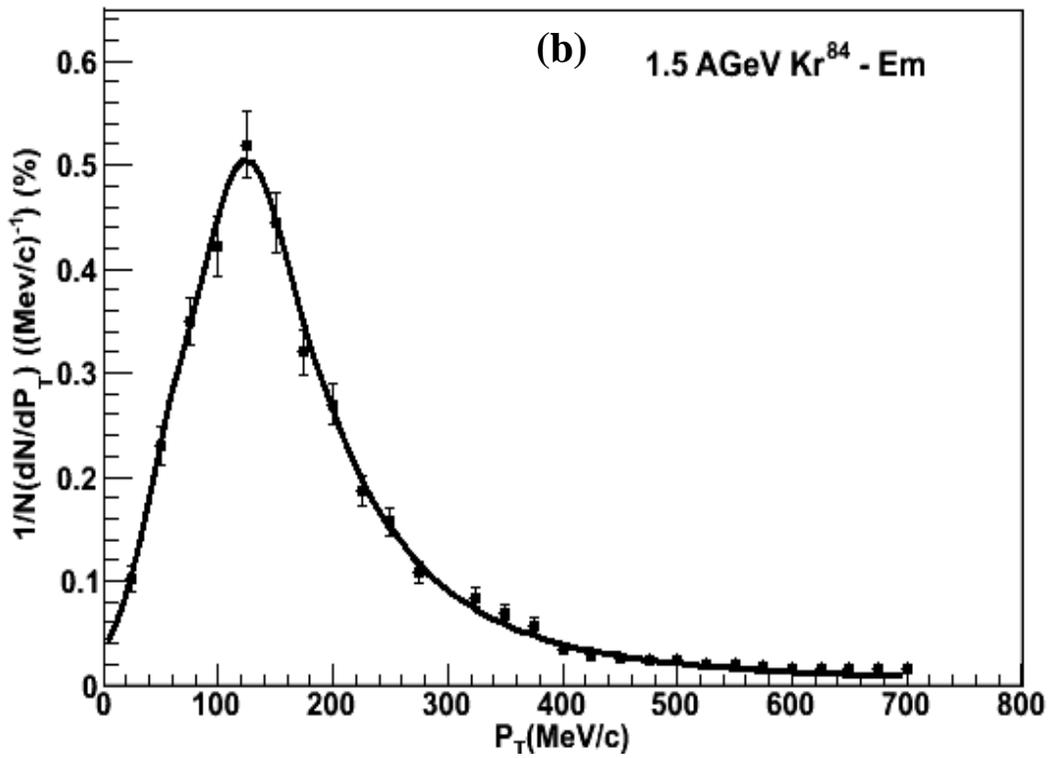

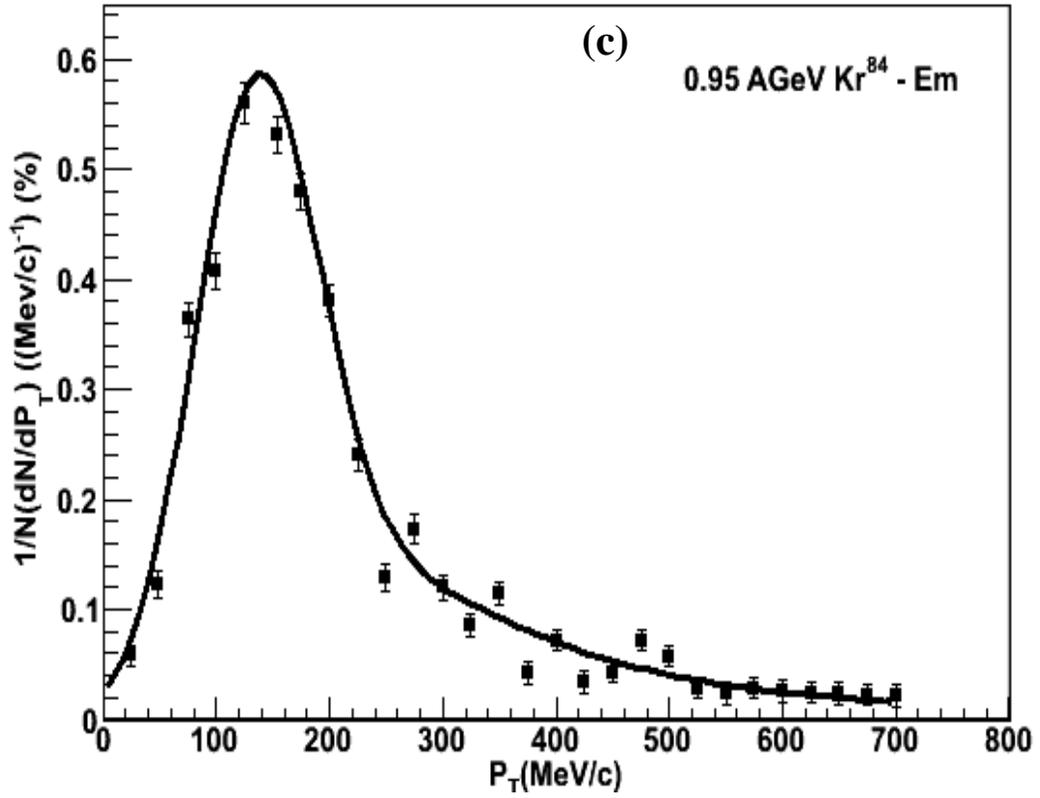

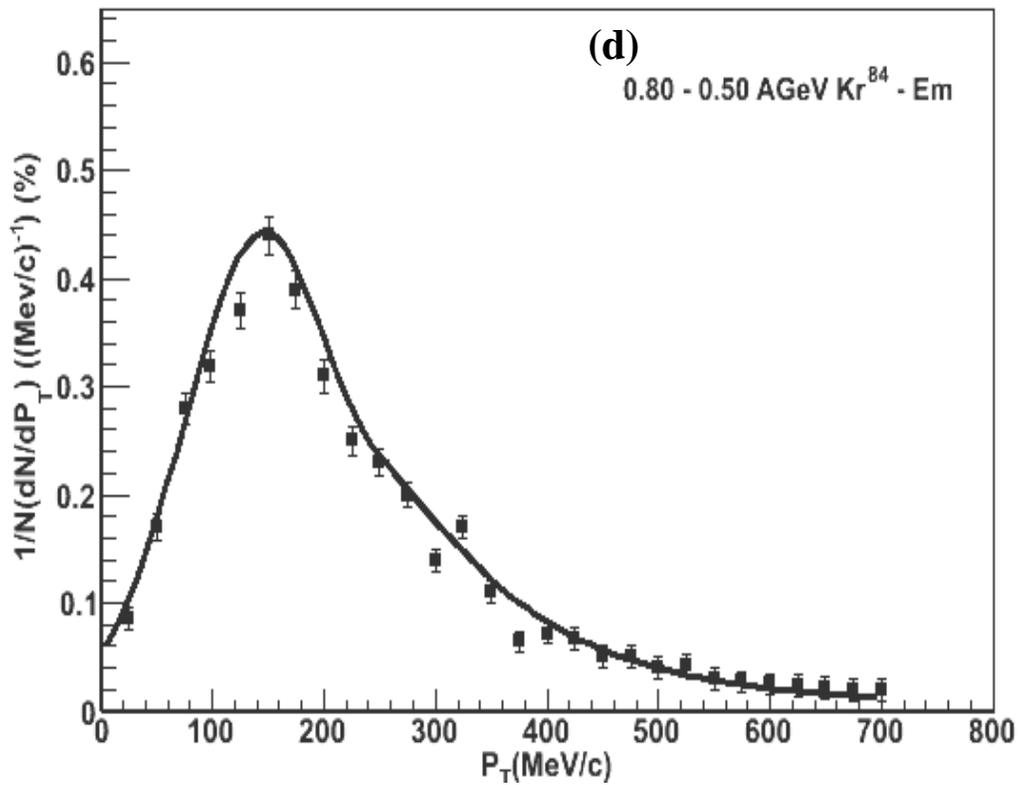

**Fig. 4:** Transverse momentum distribution of projectile fragments alpha emitted

in $^{84}$Kr nuclei interactions with emulsion target nuclei at (a) 1.7 A GeV [11], (b) 1.5 A GeV [5], (c) 0.95 A GeV and (d) 0.80-0.50 A GeV [5], respectively. The closed circles are observed values and the solid curve is the calculated values of the assumption of two source of projectile fragments alpha emission which is the sum of the two Rayleigh scattering distributions.